\documentclass[12pt]{article}
\usepackage{csvsimple}
\usepackage{graphicx, setspace, amssymb, subfigure, url, multirow, booktabs, verbatim, bm,threeparttable, amsthm, color}
\usepackage{amsmath} 
\usepackage{indentfirst}
\usepackage{bm}
\usepackage{multirow}
\usepackage[round]{natbib}
\usepackage{appendix}
\usepackage{mathrsfs}
\usepackage{graphicx}
\usepackage{subfigure}
\usepackage[ruled,vlined]{algorithm2e}
\usepackage{caption}
\usepackage{footnote}
\usepackage[figuresright]{rotating}
\usepackage{lscape}
\usepackage{float}
\usepackage{array}
\usepackage{booktabs}
\usepackage{ulem}
\usepackage{makecell}
\usepackage[colorlinks,linkcolor=blue,anchorcolor=blue,citecolor=blue,urlcolor=blue]{hyperref}
\usepackage{authblk}
\usepackage{CJKutf8}

\newtheorem{thm}{Theorem}[]

\newtheorem{con}{Condition}[]

\newtheorem{rem}{Remark}[]

\def\bt{{\boldsymbol\theta}}

\def\X{{\bf X}}

\def\wh{\widehat}
\def\wt{\widetilde}
\def\bse{\begin{eqnarray*}}
\def\ese{\end{eqnarray*}}
\def\be{\begin{eqnarray}}
\def\ee{\end{eqnarray}}
\newcommand{\bit}{\begin{itemize}}
\newcommand{\eit}{\end{itemize}}

\def\as{\mbox{ a.s.}}

\def\tr{\mbox{trace}}

\usepackage{hyperref}
\allowdisplaybreaks[3]

\def\boxit#1{\vbox{\hrule\hbox{\vrule\kern6pt  \vbox{\kern6pt#1\kern6pt}\kern6pt\vrule}\hrule}}

\begin{document}
\setlength{\textheight}{650pt}
\setlength{\baselineskip}{20pt}
\title{\bf 
	Prediction-Powered Linear Regression: A Balance Between Interpretation and Prediction}

\author{
	\ Fuzhi Xu\textsuperscript{a},
	\ Xingyu Yan\textsuperscript{b},
	and
	\ Xinyu Zhang\textsuperscript{{c}}\thanks{Correspondence address: xinyu@amss.ac.cn}\\
	\textsuperscript{a}School of Management, \\
	University of Science and Technology of China, Hefei, China \\
	\textsuperscript{b}School of Mathematics and Statistics, \\
	Jiangsu Normal University, Xuzhou, China \\
	\textsuperscript{c}Academy of Mathematics and Systems Science, \\
	Chinese Academy of Sciences, Beijing, China \\
}

\date{}
\maketitle

\begin{abstract}
	Unlabeled data are increasingly prevalent in contemporary economic studies, yet their effective use for improving prediction remains challenging because the outcomes are often costly or even infeasible to observe. Machine learning methods can help label these data and achieve high predictive accuracy, but they often lack interpretability. In this paper, we propose a Prediction-powered Unified Model Averaging (PUMA) framework to combine linear regression and machine learning methods, achieving a balance between interpretation and prediction.
	Unlike existing studies on prediction-powered inference, our approach is the first to jointly address uncertainty arising from model misspecification, power tuning parameter selection, and the choice of machine learning algorithms by using model averaging. Theoretically, under mild conditions, we establish the in-sample and out-of-sample asymptotic prediction optimality, estimation consistency, and asymptotic distribution of the PUMA estimator.
	Extensive simulations and a real-world application further demonstrate the empirical advantages of the proposed method over existing state-of-the-art approaches.
\end{abstract}


\section{Introduction}

With the rapid advancements in
data generation and acquisition technologies,
large amounts of unlabeled data
accompanied by limited labeled data are
frequently encountered
in various fields of economics, 
such as 
out-of-control risk management \citep{liu2024semi},
forecasting of economic sentiment indicators \citep{algaba2020econometrics}, quantification of intangible capital \citep{li2021measuring}, 
and financial time series analysis \citep{xi2022semi}.
For example, in development economics, a limited number of household survey observations provide the labeled poverty status for standard prediction models, while much larger census datasets contain extensive demographic information without poverty labels \citep{echevin2025combining}. 
Despite their widespread availability and
practical importance,
effectively leveraging unlabeled data remains
a fundamental challenge, both methodologically and computationally. This difficulty primarily arises because acquiring reliable labels is often expensive,
time-consuming, inconvenient,
or even infeasible.

Pre-trained machine learning (ML) algorithms provide a convenient mechanism for generating pseudo-labels in large unlabeled datasets. A natural strategy is to first use ML models to impute missing outcomes. Recently, prediction-powered inference (PPI) has emerged as a useful framework for incorporating ML-generated pseudo-labels into statistical inference procedures \citep{angelopoulos2023prediction,angelopoulos2023ppi++}. By combining labeled and pseudo-labeled data, PPI improves estimation efficiency while preserving valid statistical guarantees. It is well known that linear regression is highly interpretable, and thus we combine pre-trained ML algorithms and linear regression via PPI to achieve a balance between prediction and interpretation. However, this paradigm is highly sensitive to model uncertainty (i.e., the use of covariates) in linear regression and the quality of pre-trained ML models, which raises several fundamental concerns:

\bit

\item {\bf Model uncertainty:} 
PPI methods often select a single ``best'' model based on 
a data-driven criterion. 
This practice may result in instability: 
even small perturbations in pseudo-labels 
can lead to substantially different model choices, 
thereby affecting predictive performance.

\item {\bf Tuning parameter uncertainty:}
PPI-type estimators involve a power tuning parameter 
$\lambda \in [0,1]$  that controls the contribution of unlabeled data. 
When $\lambda=1$, the method reduces to the original PPI framework; when $\lambda=0$, it degenerates to classical supervised learning that ignores unlabeled data. 
The choice of $\lambda$ is therefore crucial yet inherently uncertain due to the quality of pseudo-labels \citep{angelopoulos2023prediction}. 
Existing theoretical selection rules often rely on complicated expressions 
and accurate estimation of asymptotic covariance matrices. 
In practice, unknown noise levels, data contamination, and outliers may severely distort variance estimation, 
leading to unstable or suboptimal tuning decisions.

\item {\bf Machine learning algorithm uncertainty:}
Different ML algorithms may exhibit substantially different predictive performance across applications. 
Treating predictions from a single selected algorithm as fixed inputs in downstream inference ignores algorithmic variability and may induce bias or inflated variance when prediction quality is inadequate \citep[e.g.,][]{wang2020methods,miao2025assumption}. 
Algorithm choice therefore introduces an additional layer of uncertainty 
that remains largely unaddressed in existing PPI frameworks.

\eit

As a result, 
jointly achieving interpretability and predictive performance remains a major practical challenge, as it requires carefully balancing model specification, tuning parameters, and ML algorithms.
To the best of our knowledge, no existing work simultaneously accounts for all these sources of uncertainty while preserving structural transparency and predictive accuracy. Therefore, a unified framework that explicitly accounts for these challenges is critically needed.

\subsection{Related Work}

{\bf\noindent
	Semi-supervised learning:} Semi-supervised learning (SSL), which leverages a small amount of labeled data together with abundant unlabeled data to improve model performance, has emerged as
a prominent paradigm in both machine learning and statistics
\citep[e.g.,][]{chakrabortty2022semi,gao2024semi,tu2024distributed,wu2025robust}.
Notably, significant progress has been made
in recent years toward understanding and
exploiting the statistical benefits of unlabeled data.
For example, \cite{chakrabortty2018efficient,li2022semi}
studied linear regression under
the SSL framework and proposed estimators
that are more efficient than ordinary least squares,
which rely solely on labeled data.
\cite{song2024general,kim2025semi} further extended this
line of work to general M-estimation problems.
These ideas have also been adapted
to high-dimensional settings \citep{zhang2022high,deng2024optimal}. 
However, these SSL methodologies have not been well integrated with advanced ML algorithms. This disconnect substantially limits the practical applicability and scalability of SSL across diverse scientific fields.

{\bf\noindent Prediction-powered inference:} More recently, a growing body of research has focused
on statistical inference
within the SSL framework by incorporating
pseudo-labels from black-box prediction models \citep[e.g.,][]{miao2025assumption,li2025prediction,xu2025unified,wang2025empirical}.
In particular, \cite{angelopoulos2023prediction} 
introduced the standard PPI framework, 
which enables valid inference in theory
even when the predictive model is of low quality.
However, although the PPI estimator provides valid confidence intervals, it can be less statistically efficient
than naive approaches that
rely solely on labeled data.
This limitation has motivated subsequent
work to improve the efficiency
of PPI or integrate it with other statistical
and ML paradigms.
Notable examples include efficient PPI (PPI++) \citep{angelopoulos2023ppi++},
cross-PPI \citep{zrnic2024cross},
stratified PPI \citep{fisch2024stratified},
and local PPI \citep{gu2024local}. These studies primarily focused on developing estimators that are asymptotically more efficient than their supervised counterparts, rather than on improving prediction performance.

{\bf\noindent Model averaging:} Model averaging is
an attractive ensemble technique for
constructing accurate predictions. Instead of relying on a single selected model, this approach aggregates information
from a collection of candidate models
by assigning them appropriate weights,
thereby reducing the risk of model misspecification compared to
single-model procedures.
In conventional regression settings,
notable examples include
optimal mean squared error
averaging \citep{hansen2007least},
jackknife model averaging \citep{hansen2012jackknife},
optimal model averaging for panel data models \citep{jiang2025robust,yu2025unified},
optimal model averaging for
generalized linear models \citep{ando2017weight,bu2025improving},
model averaging for high-dimensional
data \citep{ando2014model,wan2025high},
and quantile model averaging
\citep{lu2015jackknife,tu2025quantile},
among others.
Despite its widespread success in modern data analysis, the application of model averaging to PPI-type estimators
remains largely unexplored.
This paper fills this gap by integrating
model averaging into the prediction-powered framework,
resulting in a unified and interpretable prediction procedure,
together with formal asymptotic guarantees
for improved predictive performance.

\subsection{Our Contributions}

The proposed PUMA framework offers the following key
contributions.
\bit
\item  
To preserve interpretability, we integrate ML-generated pseudo-labels into a linear regression framework. Linear regression serves as the base model, ensuring structural transparency, while ML-predicted pseudo-labels are incorporated through the PPI++ estimator to enhance predictive accuracy. The resulting prediction-powered regression framework achieves an appropriate balance between interpretability and predictive performance.

\item  
To address the sources of uncertainty
mentioned earlier, we propose a novel model averaging strategy.  
Specifically, we aggregate a collection of explicit-form estimators constructed from candidate rectified loss functions under the linear regression framework. These candidates correspond to different combinations of models, tuning parameters, and ML algorithms, thereby explicitly accounting for all three sources of uncertainty.

\item
We develop a weight selection criterion
based on a Mallows-type measure and
determine the weights by minimizing this criterion.
This criterion serves as an unbiased estimator
of the expected in-sample squared prediction error
up to a constant.
The resulting weights lead to a prediction-driven
combination of models, tuning parameters,
and ML algorithms,
thereby further improving the overall predictive accuracy.

\item
We establish the
asymptotic optimality of the resulting model averaging prediction in both in-sample and out-of-sample settings, in the sense that its prediction risk converges to that of the infeasible oracle prediction. 
Furthermore, we prove the estimation consistency
and the asymptotic distribution of the PUMA estimator.
The presence of pseudo-labels induces a more intricate projection structure, making the analysis substantially different from that in classical supervised learning settings. 
To the best of our knowledge, our results provide the first theoretical guarantees of prediction optimality for the prediction-powered linear regression framework.

\item Our proposal is computationally efficient. 
Prediction-powered linear regression leads to a simple and easy-to-use
fitting procedure,
where the resulting estimators admit closed-form expressions.
Consequently, computation reduces to repeated linear estimation,
avoiding complex optimization routines
and substantially lowering the computational cost.
Moreover, our method does not require
access to the internal structure
of the underlying ML algorithms,
making it broadly applicable. 
\eit

\subsection{Organization}
The structure of this paper is as follows. Section \ref{Methodology} introduces the proposed PUMA method. Section \ref{sec:m:1} discusses our primary objective of balancing interpretability with predictive accuracy. Section \ref{sec:m:2} describes the adaptively combined approach and prediction-powered fitting procedure. Section \ref{sec:m:3} presents the developed weight selection criterion.
Section \ref{sec:asy} provides theoretical support for the proposed approach.
Section \ref{sec:the:1} establishes the asymptotic optimality in both in-sample and out-of-sample settings; Section \ref{sec:the:2} proves the estimation consistency; Section \ref{sec:AD} derives the asymptotic distribution of the PUMA estimator.
Section \ref{Simulation} uses Monte Carlo experiments, designed to mimic various realistic scenarios, to demonstrate the effectiveness of our approach compared to alternative methods. Section \ref{Real Data Study} illustrates the application of our method to predicting homelessness in Los Angeles. Conclusions and discussions are provided in Section \ref{Discussion}, and technical details, theoretical proofs, and additional numerical results are presented in the Supplementary Material.

\section{Methodology} \label{Methodology}

\subsection{Framework and Objective}\label{sec:m:1} 

We consider a semi-supervised setting 
with $n$ independent and identically distributed (i.i.d.) labeled observations $\{({\mathbf{X}}_i,Y_i),i=1,\ldots,n\}$ 
and $N$ i.i.d. unlabeled
observations $\{\widetilde{\mathbf{X}}_i,i=n+1,\ldots,n+N\}$, both drawn from the same underlying joint distribution. Typically, $N$ is much larger than $n$.
For the labeled data, we consider the following 
data generating process (DGP):
\begin{equation}\label{label_formula}
	\begin{aligned}
		Y_i=\mu_i + \epsilon_i = g(\mathbf{X}_i^{\ast})+\epsilon_i, \quad i=1,\ldots, n,
	\end{aligned}
\end{equation}
where 
$\mathbf{X}_i^{\ast}\in \mathbb{R}^p$ denotes the true covariate vector
in the DGP, 
$\mu_i\equiv g(\mathbf{X}_i^{\ast})$ for some unknown smooth function $g(\cdot)$, and the i.i.d. errors satisfy
$\mathbb{E}(\epsilon_i \vert \mathbf{X}_i^{\ast})=0$ and $\mathbb{E}(\epsilon_i^2 \vert \mathbf{X}_i^{\ast})=\sigma^2$.
Similarly, the unlabeled data follow the same DGP, $\widetilde{Y}_i= g(\widetilde{\mathbf{X}}_i^{\ast})+\widetilde{\epsilon}_i$, for $i=n+1,\ldots, n+N$, where the responses $\widetilde{Y}_i$ are unobserved, $\widetilde{\mathbf{X}}_i^{\ast}\in \mathbb{R}^p$ denotes the true covariate vector of the unlabeled data, $\mathbb{E}(\widetilde{\epsilon}_i \vert \widetilde{\mathbf{X}}_i^{\ast})=0$, and $\mathbb{E}(\widetilde{\epsilon}_i^2\vert\widetilde{\mathbf{X}}_i^{\ast})=\sigma^2$.
In practice, it is generally infeasible to observe all components in ${\mathbf{X}}_i^{\ast}$
or $\wt{\mathbf{X}}_i^{\ast}$, especially when $p$ tends to infinity. 
Without loss of generality, we assume that the first $q$ covariates are observed, where $q\le p$.

According to \eqref{label_formula}, the unknown DGP, together with the availability of a large amount of unlabeled data, makes it challenging to directly conduct reliable statistical modeling, construct efficient estimators, and achieve accurate predictions. To address these challenges, we propose a novel two-stage approach.

\noindent{\bf Stage I.} 
We strike a deliberate balance between interpretability and predictive accuracy. Specifically, we approximate the underlying DGP using a collection of linear working models. Although this approximation may appear simplistic, it yields a set of explicit estimators, enabling fast and stable computation as well as strong generalization performance, while simultaneously ensuring model interpretability.

\noindent{\bf Stage II.} 
We address the uncertainty inherent in semi-supervised prediction.
Specifically, we
incorporate the prediction-powered framework into our candidate linear regression models not only to leverage the unlabeled data but also to simultaneously account for uncertainty arising from model misspecification, the choice of power tuning parameters, and the selection of ML algorithms. 
Finally, we develop a data-driven weight selection criterion to unify these components into an adaptively combined procedure for accurate predictions.

Below, we present this strategy in detail. 

\subsection{ Candidate Strategies and Estimation}\label{sec:m:2}

We construct $S_1$ linear models to approximate the DGP. 
Specifically, for $s_1=1,\ldots,S_1$, the $s_1$th candidate model is 
\begin{equation}\label{submodel}
	\begin{aligned}
		Y_i= \mathbf{X}_i^{(s_1)\rm T}\bm{\theta}^{(s_1)}+e_i^{(s_1)}, \quad i=1,\ldots, n+N,
	\end{aligned}
\end{equation}
where $\bm{\theta}^{(s_1)}$ is the $s_1$th unknown coefficient vector, and, with a slight abuse of notation here, $\mathbf{X}_i^{(s_1)}$ denotes the
observed covariate vector corresponding to the $s_1$th candidate model
for the labeled data when $i=1,\ldots,n$, and the unlabeled covariates when $i=n+1,\ldots,n+N$. The term $e_i^{(s_1)}$ represents the approximation error of the $i$th observation under the $s_1$th candidate model.
Note that the unlabeled responses $\{Y_i : i=n+1,\ldots,n+N\}$ are not observed. 
Our goal is to estimate the coefficient vectors in \eqref{submodel} and
obtain reliable predictions by leveraging the information contained in the unlabeled data. To this end, we employ PPI++ \citep{angelopoulos2023ppi++} to estimate the coefficient vector for each candidate model. The detailed implementation of PPI++ is provided in the Supplementary Material.

Although PPI++ successfully exploits unlabeled data, its implementation depends on selecting an ML algorithm and a power tuning parameter, which introduces substantial uncertainty. Different algorithms perform inconsistently across datasets, the tuning parameter is highly sensitive to data perturbations, and each candidate model may require a distinct configuration. How to efficiently aggregate these heterogeneous, model-specific procedures for accurate predictions thus remains unclear.

\begin{figure}[H]
	\centering 
	\includegraphics[width=450pt]{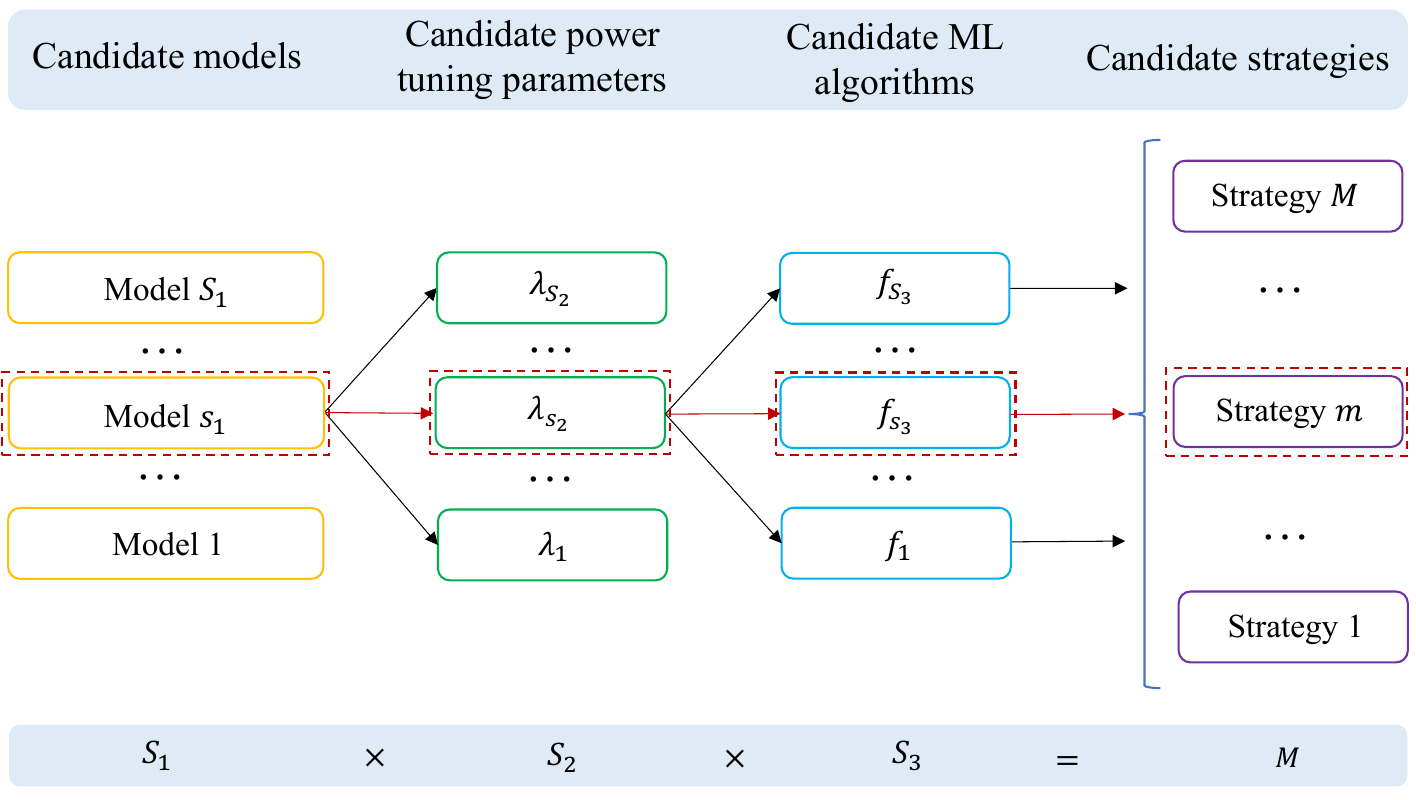}
	\caption{
		A schematic diagram of the construction of candidate strategies. 
		In addition to the candidate models, the set
		$\{\lambda_1, \lambda_2,\ldots, \lambda_{S_2} \}$ denotes the power tuning parameters, and $\{f_1,f_2, \ldots, f_{S_3} \}$ denotes the prediction functions obtained from different pre-trained ML algorithms. 
		These components are repeatedly combined across different candidate models. 
		For example, the red arrows illustrate a possible construction process of candidate strategy $m$,
		which is formed by combining the $s_1$th candidate model, the $s_2$th power tuning parameter, and the $s_3$th ML algorithm.} 
	\label{diagram}
\end{figure}

Therefore, to account for the unavoidable uncertainty arising from the choice of the power tuning parameters and pre-trained ML algorithms, 
we further 
consider $S_2$
candidate values of 
power tuning parameters and $S_3$
candidate ML algorithms for each candidate model. Consequently, by combining
the $S_1$ candidate models,
the total number of candidate strategies is $M=S_1 \times S_2 \times S_3$, where each candidate model is associated with $S_2 \times S_3$
specific selection configurations.
\footnote{
	\footnotesize 
	When zero is included in the candidate set of power tuning parameters,
	different ML algorithms may 
	yield identical candidate strategies,
	as the ML predictions $f_m(\cdot)$ are then excluded from the estimation.
	This only results in redundant configurations and does not affect
	the proposed procedure.
	If desired, these redundant strategies can be removed,
	in which case the number of distinct strategies is
	$M =S_1 \times S_2 \times S_3- S_1 \times(S_3 - 1)$.
}
For ease of exposition, the construction of all candidate strategies is illustrated in Figure \ref{diagram}. 

To simplify the notation, we reindex all candidate strategies using a unified single index 
$m$,
where $m=1,\ldots,M$.
For the $m$-th strategy, we let ${\mathbf{X}}_i^{(m)}
\in\mathbb{R}^{k_m}$ and $\widetilde{\mathbf{X}}_i^{(m)}\in\mathbb{R}^{k_m}$ denote the observed covariate vectors for the labeled data and unlabeled data, respectively.
Let
$\lambda_m$ denote the power tuning parameter, and let $ f_m: \mathbb{R}^{q}\mapsto \mathbb{R}$ denote the 
prediction function obtained
from a 
pre-trained ML algorithm. 
Note that different indices $m$ may share identical components. 
In particular, ${\mathbf{X}}_i^{(m)}$, $\widetilde{\mathbf{X}}_i^{(m)}$, $\lambda_m$, and $f_m$ may coincide across multiple $m$, and the covariates generated from the same candidate model have the same dimension. These overlaps do not affect our method and theoretical analysis, and thus we do not distinguish among such cases.
For $m=1, \ldots , M$, define $L_n^{(m)}(\bm {\theta}^{(m)})=\sum_{i=1}^n(\mathbf{X}_i^{(m)\rm T}\bm{\theta}^{(m)}-Y_i)^2/(2n)$,
$\widetilde{L}_N^{f_m}(\bm {\theta}^{(m)})=\sum_{i=n+1}^{n+N}(\widetilde{\mathbf{X}}_i^{(m)\rm T}\bm{\theta}^{(m)}-f_m(\widetilde{\mathbf{X}}_i))^2/(2N)$
and
$
{L}_n^{f_m}
( \bt^{(m)})
=\sum_{i=1}^n(\mathbf{X}_i^{(m)\rm T}\bm{\theta}^{(m)}-f_m(\mathbf{X}_i))^2/(2n)
$. 
Following the idea of PPI++, the rectified loss function for the 
$m$-th candidate strategy is defined as
\begin{align}\label{m}
	L_{\lambda_m,f_m}^{(m)}(\bm {\theta}^{(m)})
	=
	L_n^{(m)}(\bm {\theta}^{(m)}) + \lambda_m \left(
	\widetilde{L}_N^{f_m}(\bm {\theta}^{(m)})-{L}_n^{f_m}(\bm {\theta}^{(m)}) \right),
\end{align}
where 
$\lambda_m \in [0,1]$ controls the relative contribution of labeled and unlabeled data to the loss function, depending on the reliability of
$f_m$. Additional explanation 
and discussion of the rectified 
loss \eqref{m} are provided 
in the Supplementary Material A.

After a straightforward calculation, the closed-form prediction-powered solution for the $m$-th candidate is obtained by
minimizing \eqref{m}: 
\begin{align}\label{theta_hat}
	\widehat{\bm{\theta}}^{(m)}_{\lambda_m,f_m}
	= { \mathbf{\Pi}_{m} }\mathbf{\Psi}_m^{-1}\left(\mathbf{X}^{(m)\rm T}\mathbf{Y}+r{\lambda_m}\widetilde{\mathbf{X}}^{(m)\rm T}\bm{f}_m(\widetilde{\mathbf{X}})-\lambda_m\mathbf{X}^{(m)\rm T}\bm{f}_m(\mathbf{X}) \right),
\end{align}
where $\bm{f}_m(\mathbf{X})=(f_m(\mathbf{X}_1),\ldots,f_m(\mathbf{X}_n))^{\rm T}\in \mathbb{R}^n$, $\bm{f}_m(\widetilde{\mathbf{X}})=(f_m(\widetilde{\mathbf{X}}_{n+1}),\ldots,f_m(\widetilde{\mathbf{X}}_{n+N}))^{\rm T}\in \mathbb{R}^N$, and
$$\mathbf{\Psi}_m=(1-\lambda_m) \mathbf{X}^{(m)\rm T}\mathbf{X}^{(m)}+r\lambda_m\widetilde{\mathbf{X}}^{(m)\rm T}\widetilde{\mathbf{X}}^{(m)},$$
with $r=n/N$,
$\mathbf{X}^{(m)} =({\mathbf{X}}_1^{(m)},\ldots, {\mathbf{X}}_n^{(m)})^{\rm T}\in \mathbb{R}^{n\times k_m}$,
$\widetilde{\mathbf{X}}^{(m)}=(\widetilde{\mathbf{X}}_{n+1}^{(m)},\ldots,\widetilde{\mathbf{X}}_{n+N}^{(m)})^{\rm T} \in \mathbb{R}^{N\times k_m}$, $\mathbf{X}=({\mathbf{X}}_1,\ldots, {\mathbf{X}}_n)^{\rm T}\in \mathbb{R}^{n\times q}
$, 
$\widetilde{\mathbf{X}}=(\widetilde{\mathbf{X}}_{n+1},\ldots,\widetilde{\mathbf{X}}_{n+N})^{\rm T}
\in \mathbb{R}^{N \times q}
$,
and $\mathbf{Y}=(Y_1,\ldots, Y_n)^{\rm T}$. 
Here, $\mathbf{\Pi}_{m}$ is a $q\times k_m$ 
selection matrix with all elements being $1$ or $0$
such that 
$
\X{\mathbf{\Pi}}_{m}=\X^{(m)}
$ 
and 
$
\widetilde{\X}{\mathbf{\Pi}}_{m}=\widetilde{\X}^{(m)}.
$ 
By left-multiplying by the selection matrix $\mathbf{\Pi}_m$, we map the $k_m$-dimensional estimate to a $q$-dimensional space in \eqref{theta_hat}.

Let $\mathbf{w}=(w_1,\ldots,w_M)^{\rm T}$ denote the weight vector belonging to the unit simplex of $\mathbb{R}^M$, defined as $\mathcal{W}=\{\mathbf{w}\in [0,1]^{M}:\sum_{m=1}^M w_m=1 \}$.
Note that $w_m$ does not correspond to the probability that the $m$-th strategy is optimal, but rather serves as a data-driven weight reflecting its relative predictive support among the candidate strategies.
The model-averaged estimator of $\bm{\theta}$ is
then defined as 
\begin{equation}\label{ave_theta}
	\begin{aligned}
		\widehat{\bm{\theta}}(\mathbf{w}) 
		= \sum_{m=1}^M w_m \widehat{\bm{\theta}}^{(m)}_{\lambda_m,f_m}.
	\end{aligned}
\end{equation} 
Consequently,
for a new observation $\mathbf{X}_{\text{new}} \in \mathbb{R}^{q}$, the model-averaged prediction is 
\begin{equation}\label{mu_new}
	\begin{aligned}
		\widehat{\mu}_{\text{new}}(\mathbf{w}) 
		=\mathbf{X}_{\text{new}}^{\rm T}\widehat{\bm{\theta}}(\mathbf{w})=\sum_{m=1}^M w_m \mathbf{X}_{\text{new}}^{\rm T} \widehat{\bm{\theta}}^{(m)}_{\lambda_m,f_m}.
	\end{aligned}
\end{equation} 
Rather than relying on a single selected strategy, this model averaging approach leverages
information from multiple candidate strategies to account for uncertainty arising from the prediction-powered framework, thereby yielding more reliable predictions.

In the proposed prediction-powered
model averaging framework, in addition to the averaged prediction, we can also obtain the model-averaged estimator $\widehat{\bm{\theta}}(\mathbf{w})$ of the unknown coefficient vector. Unlike completely black-box approaches, the proposed method yields explicit solutions for all candidate strategies, rendering the estimation procedure both transparent and computationally efficient, which is an important consideration in modern data analysis.
Moreover, since each candidate model is a linear regression, their weighted combination preserves a linear structure, thereby maintaining the interpretability of the resulting working model.

\subsection{Weight Selection Criterion}\label{sec:m:3}

In practice, the weight vector is not observed. Based on the labeled sample, we develop a fully data-driven method for determining the weights.
Using the model-averaged coefficient vector estimator in \eqref{ave_theta}, we first construct the model-averaged estimators for 
$\bm{\mu}=(\mu_1,\ldots, \mu_n)^{\rm T}$ 
as follows. 
\begin{align}\label{eq:mu_w}
	&\widehat{\bm{\mu}}(\mathbf{w}) \nonumber\\
	={}&
	\sum_{m=1}^M w_m \mathbf{X} \widehat{\bm{\theta}}^{(m)}_{\lambda_m,f_m} \nonumber\\
	={}&  \sum_{m=1}^M w_m \left(
	\mathbf{X}^{(m)}\mathbf{\Psi}_m^{-1}\mathbf{X}^{(m)\rm T}\mathbf{Y}+r\lambda_m\mathbf{X}^{(m)}\mathbf{\Psi}_m^{-1}\widetilde{\mathbf{X}}^{(m)\rm T}\bm{f}_m(\widetilde{\mathbf{X}}) -\lambda_m\mathbf{X}^{(m)}\mathbf{\Psi}_m^{-1}\mathbf{X}^{(m)\rm T}\bm{f}_m(\mathbf{X})
	\right)\nonumber\\
	={}&  \mathbf{P}(\mathbf{w})\mathbf{Y}
	+\bm{\phi}(\mathbf{w})-\bm{\psi}(\mathbf{w}), 
\end{align}
where $\mathbf{P}(\mathbf{w})=\sum_{m=1}^M w_m\mathbf{P}^{(m)}$ with $\mathbf{P}^{(m)}=\mathbf{X}^{(m)}\mathbf{\Psi}_m^{-1}\mathbf{X}^{(m)\rm T}$,
$\bm{\phi}(\mathbf{w})=\sum_{m=1}^M w_m \bm{\phi}^{(m)}$ with 
$\bm{\phi}^{(m)}=r\lambda_m\mathbf{X}^{(m)}\mathbf{\Psi}_m^{-1}\widetilde{\mathbf{X}}^{(m)\rm T}\bm{f}_m(\widetilde{\mathbf{X}})$, and $\bm{\psi}(\mathbf{w})=\sum_{m=1}^M w_m\bm{\psi}^{(m)}$ 
with $\bm{\psi}^{(m)} 
= \lambda_m\mathbf{X}^{(m)}
\mathbf{\Psi}_m^{-1}$
$\mathbf{X}^{(m)\rm T}\bm{f}_m(\mathbf{X})$. 
Then, we define the prediction-powered Mallows-type criterion as 
\begin{equation}
	\label{C}
	\begin{aligned}
		C(\mathbf{w}) =\left\|\mathbf{Y}-\widehat{\bm{\mu}}(\mathbf{w})  \right\|^2 + 2\sigma^2\tr\left(\mathbf{P}(\mathbf{w})\right).
	\end{aligned}
\end{equation}
The optimal weight vector is obtained by
\begin{equation}\nonumber
	\begin{aligned}
		\widetilde{\mathbf{w}}= 
		\mathop{\arg\min}\limits_{\mathbf{w} \in \mathcal{W}} 
		C(\mathbf{w}).
	\end{aligned}
\end{equation}

\begin{rem}\label{rem1} Unbiasedness. Define the predictive squared loss and its corresponding risk as
	\begin{equation}\nonumber
		\begin{aligned}
			L(\mathbf{w}) = \left\|\widehat{\bm{\mu}}(\mathbf{w})-\bm{\mu}  \right\|^2~\text{and}~R(\mathbf{w}) = \mathbb{E}\left(L(\mathbf{w}) \,\middle\vert\, \mathbb{X}^{\ast} \right),		
		\end{aligned}
	\end{equation}   
	where $\mathbb{X}^{\ast}=(\mathbf{X}^{\ast\rm T},\widetilde{\mathbf{X}}^{\ast\rm T})^{\rm T}$ with 
	$\mathbf{X}^{\ast}=(\mathbf{X}^{\ast}_1,\ldots, \mathbf{X}^{\ast}_n)^{\rm T}$, and
	$\widetilde{\mathbf{X}}^{\ast}=(\widetilde{\mathbf{X}}^{\ast}_{n+1},\ldots, \widetilde{\mathbf{X}}^{\ast}_{n+N})^{\rm T}$.
	By \eqref{C}, it follows that
	\begin{align*}
		\mathbb{E}\left(C(\mathbf{w}) \,\middle\vert\, \mathbb{X}^{\ast}\right) 
		&=\mathbb{E}\left\{ \|\bm{\mu}-\widehat{\bm{\mu}}(\mathbf{w})  \|^2+\|\bm{\epsilon}\|^2+2\bm{\epsilon}^{\rm T}(\bm{\mu}-\widehat{\bm{\mu}}(\mathbf{w})) + 2\sigma^2\tr\left(\mathbf{P}(\mathbf{w})\right) 
		\,\middle\vert\, \mathbb{X}^{\ast} \right\} \\
		&=R(\mathbf{w})
		+2\mathbb{E}\left\{\bm{\epsilon}^{\rm T}(\bm{\mu}-\mathbf{P}(\mathbf{w})\mathbf{Y})
		\,\middle\vert\, \mathbb{X}^{\ast}\right\} -2\mathbb{E}\left\{\bm{\epsilon}^{\rm T}\left(\bm{\phi}(\mathbf{w})-\bm{\psi}(\mathbf{w})\right)
		\,\middle\vert\, \mathbb{X}^{\ast}\right\} \\ 
		&\quad+2\sigma^2 \tr\left( \mathbf{P}(\mathbf{w})\right)
		+n\sigma^2 \\
		&=R(\mathbf{w})  - 2\mathbb{E}\left(\bm{\epsilon}^{\rm T} \mathbf{P}(\mathbf{w})\bm{\epsilon}
		\,\middle\vert\,  \mathbb{X}^{\ast}\right)+2\sigma^2 \tr\left( \mathbf{P}(\mathbf{w})\right)
		+n\sigma^2 \\
		&=R(\mathbf{w})+n\sigma^2, 
	\end{align*}
	where $\bm{\epsilon} = (\epsilon_1, \dots, \epsilon_n)^{\rm T}$. This shows that the conditional expectation of $C(\mathbf{w})$
	differs from the risk $R(\mathbf{w})$
	only by the constant offset
	$n\sigma^2$,
	which does not depend on $\mathbf{w}$. 
	Therefore, for the purpose of selecting $\mathbf{w}$, 
	we may ignore this constant and treat $C(\mathbf{w})$
	as an unbiased surrogate for $R(\mathbf{w})$. 
	Consequently, minimizing $C(\mathbf{w})$
	yields the desired data-driven weight vector.
\end{rem}

In practice, the variance $\sigma^2$ 
in \eqref{C}
is often unknown and needs to be estimated. 
Similar to \cite{hansen2007least}, we estimate $\sigma^2$ by 
$\widehat{\sigma}^2_{M^{\ast}}=(n-k_{M^{\ast}})^{-1}\|\mathbf{Y}-\widehat{\bm{\mu}}^{(M^{\ast})}\|^2$,
where $k_{M^{\ast}}=\max\{k_1, \ldots, k_M \}$ and $\widehat{\bm{\mu}}^{(M^{\ast})}$ is obtained from the least squares estimator based on the largest approximating model fitted to the labeled data, with the power tuning parameter fixed at $\lambda_{M^{\ast}}=0$. Then the criterion in \eqref{C} can be written as 
\begin{equation}\nonumber
	\begin{aligned}
		\widehat{C}(\mathbf{w}) =\left\|\mathbf{Y}-\widehat{\bm{\mu}}(\mathbf{w})  \right\|^2 + 2\widehat{\sigma}^2_{M^{\ast}}\tr\left(\mathbf{P}(\mathbf{w})\right),
	\end{aligned}
\end{equation}
and the corresponding weight vector is 
\begin{equation}\nonumber
	\begin{aligned}
		\widehat{\mathbf{w}}= 
		\mathop{\arg\min}\limits_{\mathbf{w} \in \mathcal{W}} 
		\widehat{C}(\mathbf{w}).
	\end{aligned}
\end{equation}
Therefore, the PUMA estimators are defined as 
\begin{equation}\nonumber
	\begin{aligned}
		\widehat{\bm{\theta}}(\widehat{\mathbf{w}}) 
		=  \sum_{m=1}^M \widehat{w}_m  \widehat{\bm{\theta}}^{(m)}_{\lambda_m,f_m} 
		~\text{and}~\widehat{\bm{\mu}}(\widehat{\mathbf{w}}) 
		= \mathbf{X}\widehat{\bm{\theta}}(\widehat{\mathbf{w}}). 
	\end{aligned}
\end{equation}

\section{Theoretical Guarantees}\label{sec:asy}

In this section, we investigate the theoretical properties of the proposed PUMA procedure. We establish its asymptotic optimality for both in-sample and out-of-sample prediction, and further derive the convergence rate and asymptotic distribution of the PUMA estimator.

For notational convenience, we introduce the following definitions. 
Define the minimal risk as 
$\xi = \mathop{\inf}_{\mathbf{w} \in \mathcal{W}} R(\mathbf{w})$ and 
let $\mathbf{w}_m^{o}$ denote the $M$-dimensional weight vector whose $m$-th element is one and all others are zero.
For any symmetric matrix $\mathbf{A} \in \mathbb{R}^{p \times p}$, 
let $\kappa_{\max}(\mathbf{A})$ and $\kappa_{\min}(\mathbf{A})$ denote its largest and smallest eigenvalues, respectively. 
For a vector 
${\bm v}
=(v_1, \ldots, v_p)^{\rm T} \in \mathbb{R}^p$, define its Euclidean norm by $\|{\bm v}\|=(\sum_{j=1}^p v_j^2)^{1 / 2}$. For a matrix $\mathbf{U}=\left(u_{i j}\right) \in \mathbb{R}^{p_1 \times p_2}$, denote its $L_2$ 
norm and Frobenius norm by $ \|\mathbf{U}\|=\kappa_{\max}^{1 / 2}(\mathbf{U}^{\rm T} \mathbf{U})$
and $\|\mathbf{U}\|_{\text{F}}=(\sum_{i, j} u_{i j}^2)^{1 / 2}$, respectively.
For any positive integer $q$, 
$\mathbf{I}_{q\times q}$ denotes the $q\times q$ identity matrix, $\mathbf{0}_q$ denotes the $q$-dimensional zero vector, and  $\mathbf{0}_{q\times q}$ denotes the $q \times q$ zero matrix. Let $a \lor b$ denote $\max\{a, b\}$
and
$\mathbb I(\cdot)$ denote the indicator function.
We use 
$C_1,C_2,\ldots$,
$c_0, c_1, c_2, \ldots$ and $\widetilde{c}_0,\widetilde{c}_1, \widetilde{c}_2, \ldots$ to denote generic finite positive constants whose values may vary from line to line, and use $\as$ to indicate that a statement holds almost surely. 
Convergence in probability and convergence in distribution are 
denoted by
$\xrightarrow{p}$
and $\xrightarrow{d}$, respectively. 
Convergence of a sequence of numbers or functions is denoted by $\to$.
All limiting processes correspond to $n\rightarrow\infty$
unless stated otherwise. The proofs of all theorems are presented in the  Supplementary Material C-F.

\subsection{Asymptotic Optimality}\label{sec:the:1}

We allow both the number of candidate strategies $M$ and the number of covariates $k_m$ in each strategy to diverge. 
To facilitate the theoretical analysis, we further impose the following assumptions.

\begin{con}\label{con1} 
	There exists a fixed integer $G$ 
	such that
	(i) $\mathbb{E}(\epsilon_i^{4G}
	\vert \mathbf{X}_i^{\ast})\le \eta < \infty \as $ for all $i = 1, \dots, n$,
	where $\eta$ is a constant;
	(ii) $M \xi^{-2G}\sum_{m=1}^M\{{R}(\mathbf{w}_m^{o}) \}^{G}= {o_p(1)}$. 
\end{con}

\begin{con}\label{con2} 
	There exist constants $c_0$, $c_1$, $\widetilde{c}_0$ and $\widetilde{c}_1$, 
	such that 
	$$
	0 < c_0\le 
	\mathop{\min}_{1\le m \le M}\kappa_{\min}\left( \mathbf{X}^{(m)\rm T}\mathbf{X}^{(m)}/n\right) 
	\le\mathop{\max}_{1\le m \le M}\kappa_{\max}\left( \mathbf{X}^{(m)\rm T}\mathbf{X}^{(m)}/n\right) \le c_1 < \infty,
	$$
	$$
	0 < \widetilde{c}_0\le\mathop{\min}_{1\le m \le M}\kappa_{\min}\left(\widetilde{\mathbf{X}}^{{(m)}\rm T}\widetilde{\mathbf{X}}^{(m)}/N\right)\le  \mathop{\max}_{1\le m \le M}\kappa_{\max}\left(\widetilde{\mathbf{X}}^{{(m)}\rm T}\widetilde{\mathbf{X}}^{(m)}/N\right)\le\widetilde{c}_1 < \infty.
	$$
\end{con}

\begin{con}\label{con3}   
	(i) $\mathbb{E}(\mu_i^4)=O(1)$; 
	(ii) $n^{-1}k_{M^{\ast}}^2=O(1)$.
\end{con}

Condition \ref{con1} (i) imposes moment restrictions on the error term, which is standard in the analysis of model averaging and has been widely assumed in the existing literature; see, for example, \cite{hansen2007least,zhu2019mallows}.
Condition \ref{con1} (ii)  
imposes a relationship between the minimal strategy risk and the risks of all candidate strategies, requiring that the minimal risk $\xi$ diverge to infinity at a certain rate, so that it is not excessively smaller than the risks of the remaining strategies.
Similar conditions can be found in \cite{wan2010least}, \cite{racine2023optimal}, and the references therein.
Condition \ref{con2} imposes a mild requirement on the minimum and maximum eigenvalues of the covariate matrix. Such conditions are commonly adopted in the literature on model selection and model averaging; see, for example, \cite{wang2011gee,li2025factor}.
Condition \ref{con3} (i) restricts the magnitude of the expectation of $\bm{\mu}$, which is similar in spirit to Condition (11) of \cite{wan2010least} and Assumption 7 of \cite{li2025factor}.
Condition \ref{con3} (ii) allows the largest number of covariates $k_{M^{\ast}}$ to diverge at a
certain rate, which is standard in high-dimensional settings \citep[e.g.][]{bu2025improving,jiang2025robust}.

\begin{thm}\label{thm1} 
	Assume $\sigma^2$ is known. 
	Under Conditions \ref{con1}--\ref{con2}, we have
	\begin{equation}\label{th1_1}
		\frac{L(\widetilde{\mathbf{w}})}{\mathop{\inf}\limits_{\mathbf{w} \in \mathcal{W}}L(\mathbf{w})}
		\xrightarrow {p} 1 .
	\end{equation}
	Assume $\sigma^2$ is unknown and estimated by $\widehat{\sigma}^2_{M^{\ast}}$.
	If Conditions \ref{con1}--\ref{con3} are satisfied, then
	we have
	\begin{equation}\label{th2_1}
		\frac{L(\widehat{\mathbf{w}})}{\mathop{\inf}\limits_{\mathbf{w} \in \mathcal{W}}L(\mathbf{w})}
		\xrightarrow {p} 1.
	\end{equation}
	
\end{thm}

Equation \eqref{th1_1} of Theorem \ref{thm1} shows that the ideal PUMA constructed through the weight vector 
$\widetilde{\mathbf{w}}$ is asymptotically optimal, in the sense that its squared prediction loss converges to that of the infeasible oracle prediction, achieving the minimal possible squared loss.
Equation \eqref{th2_1} of Theorem \ref{thm1} further shows that this asymptotic optimality continues to hold when the noise variance $\sigma^2$ is unknown and is replaced by its estimator $\widehat{\sigma}^2_{M^{\ast}}$, defined in Section \ref{sec:m:3}.

We next consider the out-of-sample prediction properties.
Recall that for a new paired observation $(\mathbf{X}_{\text{new}}, Y_{\text{new}})$, the predictive squared loss is defined as 
$
L_{\text{new}}(\mathbf{w}) = \left(\widehat{{\mu}}_{\text{new}}(\mathbf{w})-{\mu}_\text{new}  \right)^2,
$
where $\widehat{{\mu}}_{\text{new}}(\mathbf{w})$ is defined in \eqref{mu_new} and ${\mu}_\text{new}$ denotes the corresponding true regression value evaluated at the
new covariate vector $\mathbf{X}_{\text{new}}^{\ast}$.
The following conditions are required.

\begin{con}\label{con4} 
	$\mathbb{E}(\|\mathbf{X}^{(m)}_i\|^4)=O(k_{M^{\ast}}^2)$ uniformly for $m=1,\ldots,M$.
\end{con}

\begin{con}\label{con5} 
	There exists a limiting value 
	$ \bt_{m}^{\star}$ such that 
	$\mathbb{E}(\| \widehat{\bm{\theta}}^{(m)}_{\lambda_m,f_m}- \bt_{m}^{\star} \|^2)
	=O(M k_m n^{-1})$, uniformly for $m=1,\ldots,M$.
\end{con}

\begin{con}\label{con6} 
	(i) $\xi/\{\mathbb{E}(\xi)\} = O_p(1)$;  (ii) $\{ (M^2k_{M^{\ast}}^{2}) \lor (n^{1/2}Mk_{M^{\ast}}^{2})\}/\{\mathbb{E}(\xi)\} = o(1)$.
\end{con}

Condition \ref{con4} imposes a mild moment condition on the covariates.
Condition \ref{con5} ensures the estimator $\widehat{\bm{\theta}}^{(m)}_{\lambda_m,f_m}$ for each candidate strategy converges to
a limit value $\bt_m^{\star}$.   
Condition \ref{con6} (i) requires that the random variation of $\xi$ not be excessively large, so that its value is primarily determined by its expectation rather than by extreme random fluctuations.
Condition \ref{con6} (ii) restricts the maximum covariate dimension $k_{M^{\ast}}$ and the number of candidate strategies $M$, and requires that the candidate strategies be sufficiently misspecified.
See \cite{zhang2016optimal} and \cite{zhang2023model} for related discussions of similar constraints.

\begin{thm}\label{thm2} 
	If Conditions \ref{con1}--\ref{con6} are satisfied, $(L(\widetilde{\mathbf{w}})-\xi)/\{\mathbb{E}(\xi)\}$ and 
	$(L(\widehat{\mathbf{w}})-\xi)/\{\mathbb{E}(\xi)\}$
	are uniformly integrable, then we have 
	\begin{equation}\label{th3_1}
		\begin{aligned}
			\frac{\mathbb{E}(L_{\text{new}}(\widetilde{\mathbf{w}}))}{\mathop{\inf}\limits_{\mathbf{w} \in \mathcal{W}}\mathbb{E}(L_{\text{new}}(\mathbf{w}))}\rightarrow 1, 
		\end{aligned}
	\end{equation}
	and
	\begin{equation}\label{th3_2}
		\begin{aligned}
			\frac{\mathbb{E}(L_{\text{new}}(\widehat{\mathbf{w}}))}{\mathop{\inf}\limits_{\mathbf{w} \in \mathcal{W}}\mathbb{E}(L_{\text{new}}(\mathbf{w}))}\rightarrow 1.
		\end{aligned}
	\end{equation}
\end{thm}
Theorem \ref{thm2} provides a new type of asymptotic optimality that
concerns the out-of-sample prediction risk and 
accounts for the randomness of $\widetilde{\mathbf{w}}$ and $\widehat{\mathbf{w}}$.
It shows that the proposed PUMA is asymptotically optimal in the sense of attaining the lowest possible out-of-sample prediction squared error, a necessary yet insufficiently studied property in prediction theory
\citep[e.g.,][]{hansen2012jackknife,zhang2023model}.

\subsection{Estimation Consistency} \label{sec:the:2}

We will show that when there exists at least one 
correctly specified model, the PUMA estimator attains the same convergence rate as that of the correctly specified model,
provided that the DGP follows the same linear
regression framework as the candidate models.
Without loss of generality, we assume that the working model corresponding to the $m^\ast$th candidate strategy is correctly specified; that is, 
$\mu_i=\mathbf{X}_i^{\rm T} \bt_{m^\ast}^{\star}$
for $i=1,\ldots,n$.

\begin{thm}\label{thm3}
	If Conditions \ref{con2} and \ref{con5} are satisfied, then we have
	\be\label{th4_1}			
	\big\|\wh{\bt}(\widetilde{\mathbf{w}})
	-\bm\theta\big\|=O_p\big(\sqrt{{Mk_{M^{\ast}}}/{n}}\big),
	\ee
	and
	\be\label{th4_2}		
	\big\|\wh{\bt}(\widehat{\mathbf{w}})
	-\bm\theta\big\|=
	O_p\big(\sqrt{{Mk_{M^{\ast}}}/{n}}\big).
	\ee
\end{thm}

Theorem \ref{thm3}
establishes the $\sqrt{{Mk_{M^{\ast}}}/{n}}$-consistency of the PUMA estimator 
when the correctly specified models are contained in the candidate strategy set.
Note that the rate in Theorem \ref{thm3} relies on the uniform convergence rate specified in Condition \ref{con5}.
If $M$ is a fixed constant in Condition \ref{con5},
then the results in \eqref{th4_1} and \eqref{th4_2} both reduce to $O_p(\sqrt{k_{M^{\ast}}/n})$.
Furthermore, if $k_{M^{\ast}}$ is also a fixed constant, this rate reduces to $O_p(1/\sqrt{n})$, which coincides with the classical convergence rate for a parametric estimator.

\subsection{Asymptotic Distribution}\label{sec:AD}

In this section, we establish the asymptotic distribution of the PUMA
estimator when both $M$ and $q$ are fixed.
Let $\mu_i=\mathbf{X}_i^{\rm T}\bm{\theta}=
\mathbf{X}_{1i}^{\rm T}\bm{\beta}+\mathbf{X}_{2i}^{\rm T}\bm{\gamma}$ for the labeled data, where $\mathbf{X}_i=(\mathbf{X}_{1i}^{\rm T},\mathbf{X}_{2i}^{\rm T})^{\rm T} \in \mathbb{R}^q$ and $\bm{\theta}=(\bm{\beta}^{\rm T},\bm{\gamma}^{\rm T})^{\rm T}\in \mathbb{R}^q$.
Similarly, for the unlabeled data, 
$\widetilde{\mu}_i=\widetilde{\mathbf{X}}_i^{\rm T}\bm{\theta}=
\widetilde{\mathbf{X}}_{1i}^{\rm T}\bm{\beta}+\widetilde{\mathbf{X}}_{2i}^{\rm T}\bm{\gamma}$.
Without loss of generality, we assume that $\mathbf{X}_{1i}$ and $\widetilde{\mathbf{X}}_{1i}$ are the core covariates, which are included in each candidate strategy, 
while $\mathbf{X}_{2i}$ and $\widetilde{\mathbf{X}}_{2i}$ are the auxiliary covariates that may or may not be included.  Accordingly, for the $m$-th candidate strategy, $\mathbf{X}^{(m)} \in \mathbb{R}^{n\times k_m}$ and $\widetilde{\mathbf{X}}^{(m)}\in \mathbb{R}^{N\times k_m}$
denote the corresponding design matrices for the labeled and unlabeled datasets, respectively, consisting of all core covariates together with a subset of the auxiliary covariates.
We impose the following regularity conditions.

\begin{con}\label{con7} 
	$\bm{\gamma} = \bm\delta/\sqrt{n}$, where $\bm\delta$ is an unknown constant vector.
\end{con}

\begin{con}\label{con8} 
	$\mathbf{X}^{\rm T}\mathbf{X}/n \xrightarrow {p} \mathbf{Q}$ and 
	$\widetilde{\mathbf{X}}^{\rm T}\widetilde{\mathbf{X}}/N
	\xrightarrow {p} \mathbf{Q}$, 
	where $\mathbf{Q}=\mathbb{E}(\mathbf{X}_i\mathbf{X}_i^{\rm T})\in \mathbb{R}^{q \times q}$ is a positive definite matrix.
\end{con}

\begin{con}\label{con9}
	(i) $\mathbb{E}(\|\mathbf{X}_i \|^4)=O(1)$; (ii) The prediction function $f_m(\cdot)$ is uniformly bounded; that is, $
	\max_{ 1\le m\le M}
	\sup_{\mathbf{x} \in \mathbb{R}^q}\vert f_m(\mathbf{x})\vert
	\le C_f<\infty$, where $C_f$ is a positive constant independent of $m$ and $\mathbf{x}$.
\end{con}

Condition \ref{con7} requires that the partial correlations between the auxiliary covariates and the response are weak,  implying that all candidate submodels become asymptotically close to each other as the sample size increases. The same condition was used in \citet{hjort2003frequentist} and \cite{liu2015distribution}.
Condition \ref{con8} is a standard regularity condition in the context of regression analysis; see, for example, \cite{zhang2019inference}
and
\cite{song2026bootstrap}.
Condition \ref{con9} (i) is a standard moment condition.
When the candidate model containing all observed covariates has a fixed 
$q$, this condition reduces to a special case of Condition \ref{con4}. 
Condition \ref{con9} (ii) places no restriction 
on the predictive accuracy of the machine learning algorithms. Instead, it merely rules out unbounded predictions by assuming that
$f_m(\cdot)$ is uniformly bounded.
This technical condition is needed to establish the asymptotic distribution in Theorem \ref{thm4}. Similar boundedness assumptions are commonly imposed in the nonparametric and machine learning literature to facilitate asymptotic analysis \citep[e.g.,][]{hastie2009elements,yan2025deep}.

\begin{thm}\label{thm4} 
	Suppose that Condition \ref{con1} (i), Condition \ref{con4} and Conditions \ref{con7}--\ref{con9} hold.  Let
	$\mathbf{v}^{\ast} = \arg\min_{\mathbf{w} \in \mathcal{W}} \mathbf{w}^{\rm T} \mathbf{\Delta}^{\ast} \mathbf{w}$
	be the unique minimizer
	over $\mathcal{W}$\as. Then,
	$\widehat{\mathbf{w}} \xrightarrow{d} \mathbf{v}^{\ast},$
	and 	
	\begin{equation}\nonumber
		\sqrt{n}\bigl(\widehat{\bm{\theta}}(\widehat{\mathbf{w}}) - \bm{\theta}\bigr) \xrightarrow{d} \sum_{m=1}^M v_m^{\ast} \bm{\Lambda}_m,
	\end{equation}
	where $\bm{\Lambda}_m=\mathbf{U}_m \widetilde{\mathbf{Z}}_m + (\mathbf{U}_m \mathbf{Q} - \mathbf{I}_{q\times q})\bm{\eta}_m$ with $\mathbf{U}_m=\mathbf{\Pi}_m\mathbf{Q}_m^{-1} \mathbf{\Pi}_m^{\rm T}  \in \mathbb{R}^{q \times q}$,
	$ \mathbf{Q}_m = \mathbb{E}(\mathbf{X}_i^{(m)} \mathbf{X}_i^{(m){\rm T}}) \in \mathbb{R}^{k_m \times k_m}$,
	and $\bm{\eta}_m = \sqrt{n}( \mathbf{I}_{q\times q}- \mathbf{\Pi}_m \mathbf{\Pi}_m^{\rm T})\bm{\theta}\in \mathbb{R}^{q}$. Here, ${v}_m^{\ast}$ denotes the $m$-th element of 
	$\mathbf{v}^{\ast}\in\mathbb{R}^M$ and $\mathbf{\Delta}^{\ast}$ is an $M\times M$ matrix of which the $(m,m')$th element is $\Delta_{m,m'}^{\ast}=-\mathbf{Z}_\epsilon^{\rm T} \bm{\Lambda}_{m'} - \bm{\Lambda}_{m}^{\rm T} \mathbf{Z}_\epsilon + \bm{\Lambda}_{m}^{\rm T} \mathbf{Q} \bm{\Lambda}_{m'} + \sigma^2\{\tr(\mathbf{U}_{m}\mathbf{Q}) + \tr(\mathbf{U}_{m'}\mathbf{Q})\}$. The definitions of $\widetilde{\mathbf{Z}}_m$ and $\mathbf{Z}_\epsilon$ are given in Supplementary Material B.
\end{thm}

Theorem \ref{thm4} shows that the PUMA estimator admits an asymptotic representation as a nonlinear function of Gaussian random vectors. This nonstandard asymptotic behavior is induced by the data-driven weight estimation in the PUMA procedure. 

\section{Simulation}\label{Simulation}

\subsection{Simulation Designs}

Let the true covariates in the DGP be denoted by
$\mathbf{X}_i^{\ast}=(\mathbf{X}_{i1}^{\ast}, \ldots, \mathbf{X}_{ip}^{\ast})^{\rm T}$, $i=1,\ldots, n$, and $\widetilde{\mathbf{X}}_i^{\ast}=(\widetilde{\mathbf{X}}_{i1}^{\ast}, \ldots, \widetilde{\mathbf{X}}_{ip}^{\ast})^{\rm T}$, $i=n+1,\ldots, n+N$, 
where both 
$\mathbf{X}_i^{\ast}$ and $\widetilde{\mathbf{X}}_i^{\ast}$ are 
independently sampled from a multivariate normal distribution
$\mathcal{N} (\bm{0}_p,\bm{\Sigma})$.
The covariance matrix is specified as $\bm{\Sigma}=\{\sigma_{ij}\}_{i,j=1}^p$ with $\sigma_{ij}=\rho^{\vert i-j\vert}$, where $\rho=0.5$
and $p=6$. The true coefficient vector is set to 
$\bm{\theta}=(1, -1.1, 0.2, -0.025, 0, \alpha)^{\rm T}$, where $\alpha$ takes values $0.1$ or $0$. 
The labeled responses are generated according to
$Y_{i}= \mathbf{X}_{i}^{\ast\rm T} \bm{\theta}+\epsilon_{i}$, where $\epsilon_{i} \sim \mathcal{N} (0,\sigma^2_{\epsilon})$ for $i=1,\ldots, n$.
We consider labeled sample sizes 
$n=50$ and $n=100$, 
with the unlabeled sample size fixed at
$N=500$. 
This setting mimics practical scenarios where unlabeled observations are abundant while labeled data are relatively limited.

Suppose the first $p-1$ covariates are observed (i.e., $q=p-1$).
To simplify computations and without loss of generality, 
we use them to form a sequence of nested candidate models
(i.e., $S_1=5$).
Specifically, based on the labeled data, we fit a linear regression using these 
$q$ covariates, compute the marginal
$p$-value for each covariate, and then build nested candidate models by adding covariates in ascending order of their 
$p$-values.
When $\alpha=0.1$, all candidate models are misspecified; when $\alpha=0$, the correctly specified model is included among the candidates and coincides with the largest model.
For each candidate model,
we consider $S_2=5$ candidate power tuning parameters, chosen from the set $\{0, 0.25, 0.5, 0.75, 1\}$.

Following \cite{angelopoulos2023ppi++},
for all $m=1,\ldots,M$,
we generate pseudo-labels $\{f_m(\mathbf{X}_i):i=1,\ldots,n \}$ and $\{f_m(\widetilde{\mathbf{X}}_i):i=n+1,\ldots,n+N \}$ by adding
noise $\varepsilon_{i}^{(m)}$
and $\widetilde{\varepsilon}_{i}^{(m)}$ to the linear predictors $\mathbf{X}_i^{\ast\rm T}\bm{\theta}$ and $\widetilde{\mathbf{X}}_{i}^{\ast\rm T}\bm{\theta}$, respectively, where $\varepsilon_{i}^{(m)}, \widetilde{\varepsilon}_{i}^{(m)}\sim \mathcal{N}(\mu_{\varepsilon},\sigma_{\varepsilon}^{2})$.
We consider $S_3=2$ ML algorithms, denoted by ML1 and ML2, and examine 
two cases. 
In Case I, 
ML2 
performs better than
ML1,
with 
$\text{ML1}:(\mu_{\varepsilon},\sigma_{\varepsilon})= (1, 0.75)$ and 
$\text{ML2}:(\mu_{\varepsilon},\sigma_{\varepsilon})= (-0.5, 0.5)$. 
In Case II, 
ML1 
performs better than
ML2,
with
$\text{ML1}: (\mu_{\varepsilon},\sigma_{\varepsilon})= (0.5, 0.25)$
and $\text{ML2}: (\mu_{\varepsilon},\sigma_{\varepsilon})=(-0.5, 0.5)$. 
Overall, the candidate ML algorithms perform better in Case II than those in Case I. 
In this simulation setting, we consider a total of $5\times 5\times2-5\times(2-1)=45$ candidate strategies.

We assess the estimators using their out-of-sample prediction performance.
Specifically, we generate an independent testing dataset $\{(\mathbf{X}_{i}^{\text{test}}, Y_{i}^{\text{test}}): i=1,\ldots,n_\text{test}\}$ with sample size $n_{\text{test}}=n$, matching the size of the labeled data.
The out-of-sample mean squared error (OMSE) is computed as
$$\text{OMSE}=
\frac{1}{n_{\text{test}}}
\|\widehat{\bm{\mu}}_{\text{test}}(\widehat{\mathbf{w}})-\bm{\mu}_{\text{test}}\|^2,$$
where $\widehat{\bm{\mu}}_{\text{test}}(\widehat{\mathbf{w}})$
is obtained from \eqref{mu_new} using the test covariate vector $\mathbf{X}_{i}^{\text{test}}$ and $\bm{\mu}_{\text{test}}$ is the vector of corresponding true regression means.
To further examine the robustness of the proposed method under varying noise levels, we define $R^2=\text{Var}(\mathbf{X}_i^{\ast\rm T}\bm{\theta})/{\text{Var}(Y_i)}$ and let $R^2$ vary from $0.1$ to $0.9$ in increments of $0.1$ by adjusting $\sigma_{\epsilon}^2$. For ease of comparison, the OMSE from each alternative method is normalized by the OMSE of the proposed method, and the resulting relative OMSE values are reported. All results are averaged over $500$ simulation replications.

\subsection{Alternative Methods}\label{sec:sim:alter}
For comparison, in addition to our PUMA method, we consider the following alternative methods: 
(a) 
Prediction-powered equally weighted model averaging (PEMA). This method follows the same procedure as the proposed approach, except that equal weights are assigned to all candidate models.
(b) 
Prediction-powered Akaike information criterion model selection (PAIC).
This approach computes the AIC value, 
$\text{AIC}_m =n \log(\widehat{\sigma}_m^2)+2\tr(\mathbf{P}^{(m)})$,
for all $m=1, \ldots,M$, where $\widehat{\sigma}_m^2=n^{-1}\|\mathbf{Y}-\mathbf{X}\widehat{\bm{\theta}}^{(m)}_{\lambda_m,f_m} \|^2$, $\widehat{\bm{\theta}}^{(m)}_{\lambda_m,f_m} $ is defined in \eqref{theta_hat}, and $\mathbf{P}^{(m)}$ is given in \eqref{eq:mu_w}. Finally, the optimal candidate strategy
is selected by minimizing $\text{AIC}_m$ over $m=1,\ldots,M$.
(c) 
Prediction-powered Bayesian information criterion model selection (PBIC). 
This approach follows the same principle as PAIC but replaces the AIC
with the BIC, $\text{BIC}_m =n \log(\widehat{\sigma}_m^2)+\tr(\mathbf{P}^{(m)})\log(n)$, and selects the strategy with the smallest $\text{BIC}_m$ for $m=1,\ldots,M$.
(d)
Prediction-powered model averaging under the largest model (PLARM). 
This method follows the same procedure as the proposed approach, except that the prediction-powered estimation is based on the largest model including all observed covariates. 
As a result, it ignores model uncertainty. 
(e) 
Prediction-powered model averaging with the first ML algorithm (PML1). This method follows the same procedure as the proposed approach, except that only ML1 is considered. 
Consequently, it ignores the uncertainty arising from the choice of ML algorithms. 
(f) 
Prediction-powered model averaging with the second ML algorithm (PML2). This approach follows the same principle as PML1, but uses ML2 instead of ML1.
(g) 
Prediction-powered model averaging with power tuning $\lambda_m=1$ (PLAM1).
This method follows the same procedure as the proposed approach, except that all candidate power tuning parameters are fixed at $\lambda_m=1$
for $m=1,\ldots,M$. Note that, for each candidate model and algorithm, this approach degenerates to the recently proposed PPI method
\citep{angelopoulos2023prediction}.  
(h) 
Prediction-powered model averaging with power tuning 
$\lambda_m=0$ (PLAM0). 
This approach also adopts our model averaging strategy to address model uncertainty, but relies solely on the labeled data. 
(i) 
The 
PPI++ method in \cite{angelopoulos2023ppi++} uses the largest model with ML1 (PPI++ML1).
(j) 
The corresponding PPI++ method uses the largest model with ML2 (PPI++ML2). 
Neither method accounts for uncertainty arising from power tuning. 
Instead, the optimal power tuning is selected via the plug-in estimator given in Equation (8) of \cite{angelopoulos2023ppi++}.
(k) 
Largest model (LARM). This method adopts the largest model and is fitted exclusively using the labeled data.
Note that both PLAM0 and LARM ignore the unlabeled information. Moreover, among all the considered methods, only PUMA and PEMA simultaneously account for uncertainty arising from model specification, tuning parameters, and ML algorithms.

\subsection{Simulation Results}

\begin{figure}[H]
	\centering 
	\includegraphics[width=490pt]{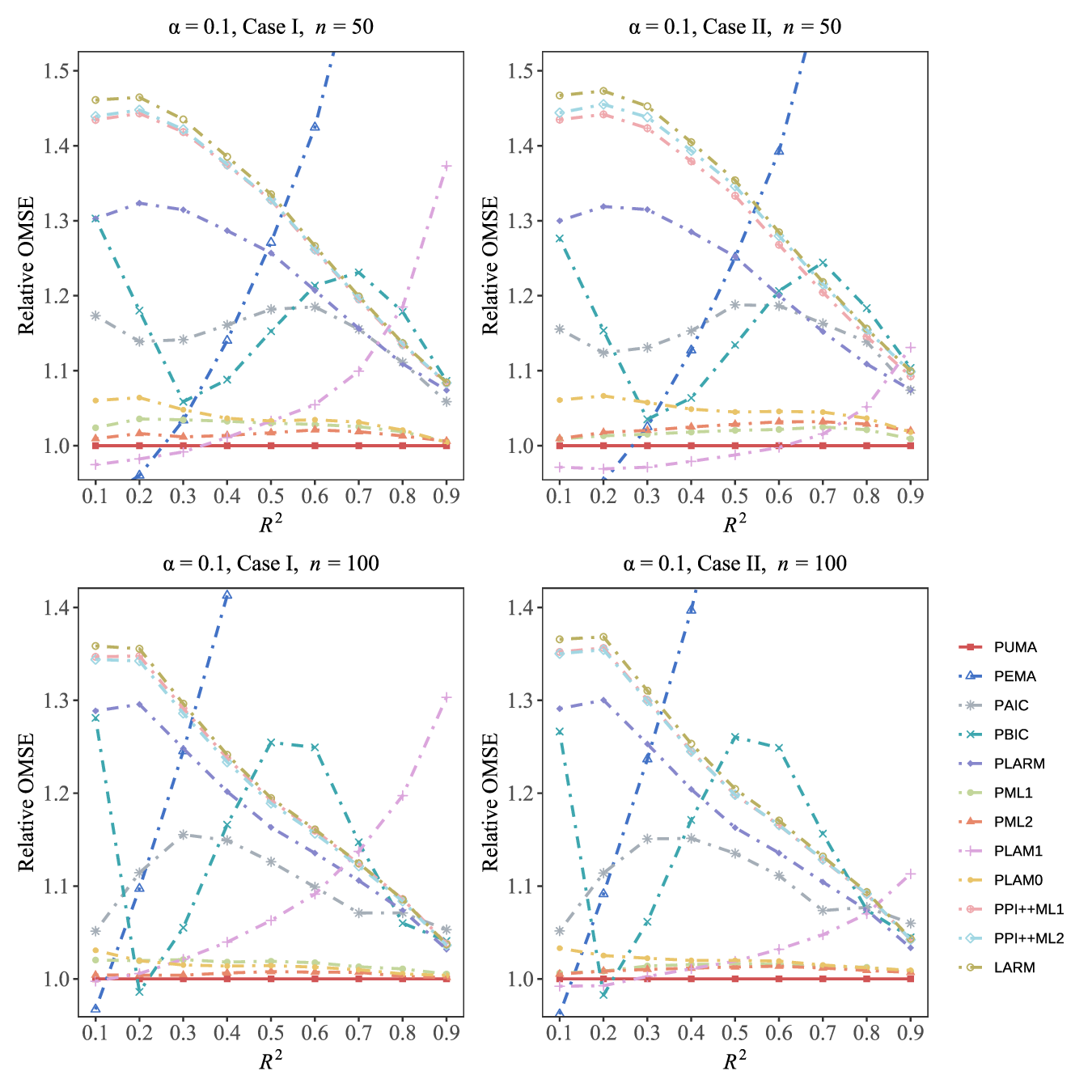}
	\caption{
							Relative OMSE for labeled sample sizes $n=50$ and $100$ with varying $R^2$, based on $500$ Monte Carlo replicates, under Cases I and II when all candidate models are misspecified (i.e., $\alpha=0.1$).		} 
	\label{R_mis_OMSE}
\end{figure}

\begin{figure}[H]
	\centering 
	\includegraphics[width=490pt]{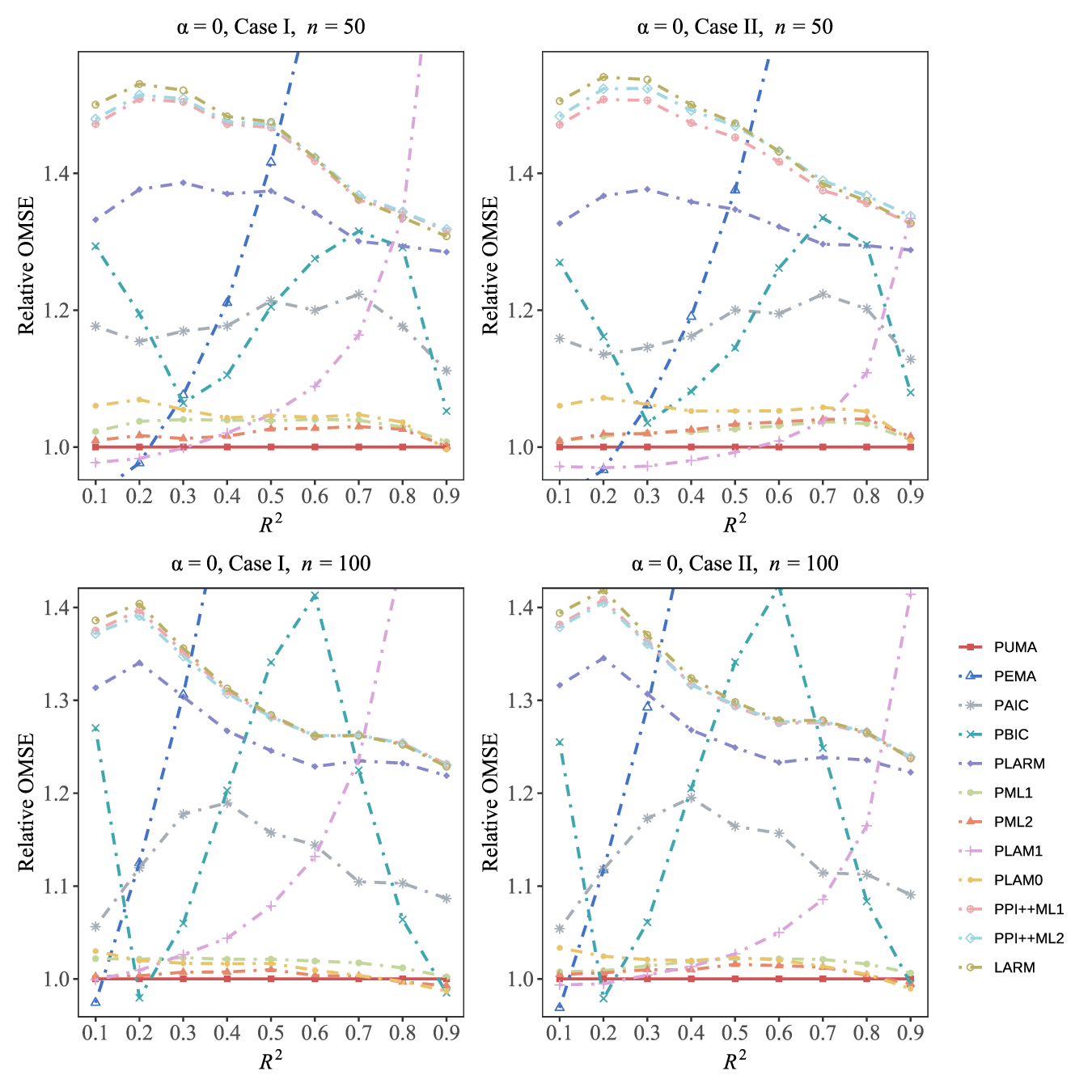}
	\caption{Relative OMSE for labeled sample sizes $n=50$ and $100$ with varying $R^2$, based on $500$ Monte Carlo replicates, under Cases I and II when the correctly specified model is included among the candidates (i.e., $\alpha=0$). } 
	\label{R_OMSE}
\end{figure}

\begin{figure}[H]
	\centering 
	\includegraphics[width=490pt]{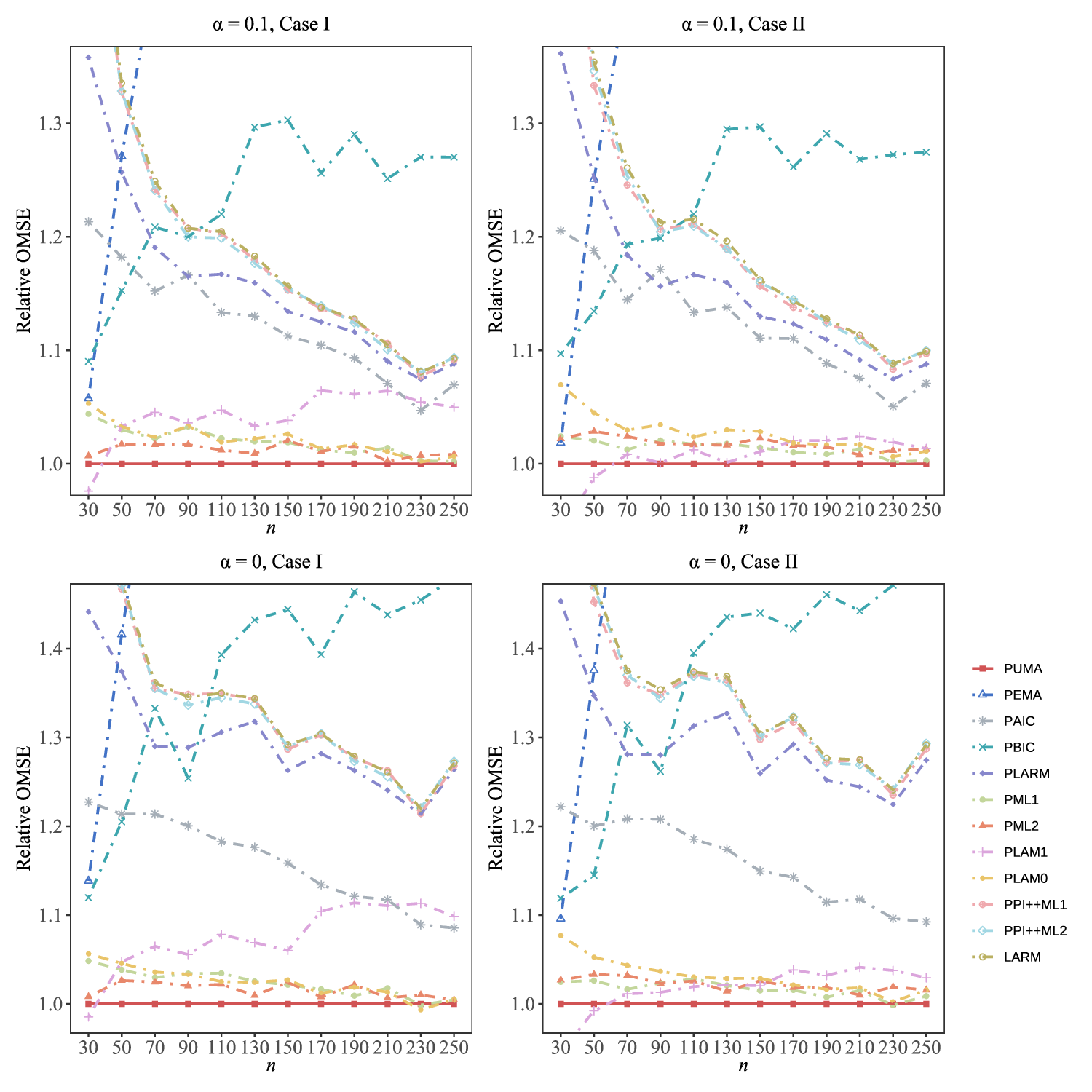}
	\caption{ Relative OMSE for $R^2=0.5$ with varying $n$, based on $500$ Monte Carlo replicates, under Cases I and II with $\alpha=0.1$ and $\alpha=0$, respectively.} 
	\label{OMSE_n}
\end{figure}

The simulation results for out-of-sample prediction are presented in Figures \ref{R_mis_OMSE}--\ref{OMSE_n}. In particular,
Figures \ref{R_mis_OMSE} and \ref{R_OMSE} summarize the relative OMSE under labeled sample sizes  $n=50, 100$ and
coefficient settings
$\alpha=0.1,0$,
respectively,
across varying values of $R^2$. 
It is clear that the proposed method achieves competitive out-of-sample prediction performance and provides substantial improvements in prediction accuracy compared with alternative methods in most settings. 
The corresponding in-sample prediction results, which exhibit similar patterns, are provided in Figures G1--G3 in the Supplementary Material G.

As shown in Figure \ref{R_mis_OMSE}, under Case I, regardless of whether the labeled sample size is small (i.e., $n=50$) or large (i.e., $n=100$), all alternative methods yield relative OMSE values greater than or equal to 1 over nearly the entire range of
$R^2$
when all candidate models are misspecified (i.e., $\alpha=0.1$). This indicates that the proposed method consistently achieves superior out-of-sample prediction accuracy.  Moreover, several alternative methods of model selection, including PAIC and PBIC, PLARM, PPI++ML1, PPI++ML2 and LARM,  
exhibit less stable performance, with their relative OMSEs remaining relatively high or fluctuating substantially across different settings. This suggests that methods based on selecting a single model may be more sensitive to model specification uncertainty.
We also observe that the PEMA method exhibits a particularly sharp increase in OMSE as 
$R^2$ increases, indicating that simple model averaging with equal weights performs substantially worse than the proposed approach. This finding underscores the advantage of the adaptive weighting mechanism of PUMA. 
In addition, the alternative methods, including PML1, PML2, PLAM1 and PLAM0, which ignore either the uncertainty arising from ML algorithms or the uncertainty associated with the power tuning parameters, also exhibit inferior predictive performance compared with our proposed method. This further highlights the necessity of incorporating these two types of uncertainty
into a prediction-powered mechanism.

Furthermore, under Case II, the proposed method also demonstrates superior out-of-sample predictive performance, and the overall trend of relative OMSE for the alternative methods as $R^2$ increases is similar to that observed in Case I.
Notably, the performance of PLAM1 in Case II is substantially better than that in Case I, which is attributed to the higher label prediction accuracy of the ML algorithms in Case II.
When the labeled sample size is $n=50$, 
PLAM1 attains lower OMSE values than the proposed method for small $R^2$. 
However, as the labeled sample size increases, this advantage of PLAM1 diminishes. Consequently, the proposed method ultimately achieves dominant out-of-sample prediction performance compared with PLAM1 and all other alternative methods.
Moreover, we observe that PML2 performs slightly better than PML1 under Case I,  whereas the difference is not significant and even exhibits the opposite pattern in Case II. This is because ML2 achieves higher predictive accuracy than ML1 in Case I, whereas ML1 is slightly more accurate in Case II. 
This observation suggests that different ML algorithms may yield markedly different predictive performance. Consequently, in addition to model and tuning uncertainty, accounting for the uncertainty in ML algorithms is essential.
In addition, the proposed method exhibits similarly strong performance in
Figure \ref{R_OMSE} when the candidate set contains a correctly specified model (i.e., $\alpha=0$), 
and we therefore omit a detailed discussion. These results further demonstrate that PUMA can effectively leverage the information contained in the unlabeled data.

Next, we explore the effect of different labeled sample sizes $n$ with fixed $R^2=0.5$. 
The results are presented in 
Figure \ref{OMSE_n} with $n$ increasing gradually from 30 to 250. 
As we can see, the proposed method consistently achieves the lowest OMSE among all alternative methods for most values of $n$.
Specifically, when $n$ is small, PUMA clearly outperforms PLAM0. As $n$ continues to increase, the relative OMSE of PLAM0 gradually approaches that of PUMA, yet it remains uniformly worse. This behavior aligns well with practical intuition: when labeled data are scarce, effectively leveraging information from unlabeled data can substantially improve predictive performance.
Comparing Case I and Case II, we observe that PLAM1 achieves better predictive performance for small sample sizes under Case II, whereas this advantage diminishes noticeably under Case I. This shows that PLAM1, which fully trusts the ML predicted labels, is unstable and highly sensitive to the prediction accuracy of the underlying algorithm.
These results also clearly demonstrate that it is necessary to account for the uncertainty in power tuning parameters, especially when the ML algorithms fail to consistently maintain high predictive accuracy.
The remaining alternative methods, including PEMA, PAIC, PBIC, PLARM, PML1, PML2, PPI++ML1, PPI++ML2, and LARM, all show substantially higher relative OMSE than PUMA across all cases. These results confirm that, in finite-sample settings, the proposed method achieves superior out-of-sample prediction performance across a wide range of labeled sample sizes.

\section{
	Empirical Application: Predicting the Homeless
}\label{Real Data Study}

In this section, we apply the proposed PUMA method to the Los Angeles homeless dataset to examine its practical usefulness. The dataset is obtained from a survey of the homeless population in Los Angeles County conducted by the Los Angeles Homeless Services Authority (LAHSA) during 2004--2005. To reduce survey costs, LAHSA adopted a stratified sampling design. Specifically, 244 areas that were considered to have a large homeless population were preselected and fully surveyed (referred to as hotspots). For the remaining areas 
(non-hotspots), 265 regions were randomly selected and visited, while the remaining 1,545 areas were not surveyed. In addition to the homeless counts, several covariates were available for all 2,054 areas.
From a semi-supervised learning perspective, the labeled and unlabeled data are drawn from the 1,810 non-hotspot areas, with sample sizes of 265 and 1,545, respectively. 
An important practical objective is to predict the number of homeless individuals in a given region, which is a significant public policy concern.

As suggested by \cite{zhang2019semi},
we consider 
seven covariates in total, 
including the percentage of land used for commercial purposes
(\textit{Commercial}), the percentage of land used for industrial purposes
(\textit{Industrial}), the median household income (\textit{MedianHouseholdIncome}),
the percentage of population that is non-Caucasian (\textit{PctMinority}),
the percentage of owner-occupied housing units (\textit{PctOwnerOcc}),
the percentage of unoccupied housing units (\textit{PctVacant}), and the 
percentage of land used for residential purposes (\textit{Residential}).
The description of these covariates is provided in Table H1 of Supplementary Material H, and more detailed information can be found in \cite{kriegler2010small}.

We next describe the construction of the candidate strategies. (a) 
Candidate models. Following the nested candidate model approach \citep[e.g.,][]{jiang2025robust}, 
the first candidate model 
includes
\textit{Commercial}, \textit{Industrial}, and \textit{Residential}, which represent fundamental spatial and economic characteristics.
The remaining candidate models are constructed sequentially by adding one covariate at a time according to the $p$-value ranking. Specifically, \textit{PctVacant}, \textit{PctOwnerOcc}, \textit{MedianHouseholdIncome}, and \textit{PctMinority} are added in this order to form candidate models two through five.
Details of these candidate models are provided in (H.1)
of Supplementary Material H. 
(b) 
Candidate ML algorithms. Two widely used ML algorithms are considered: XGBoost (ML1) and random forest (ML2), which are pre-trained
on the 244 hotspot areas described in the previous paragraph
\citep{gan2024prediction}.
(c) 
Candidate tuning parameters.
Consistent with the numerical simulation, five power tuning parameters $\{0, 0.25, 0.5, 0.75, 1\}$ are considered. Consequently, a total of 45 candidate strategies are evaluated in this real data analysis.

We randomly split the 265 labeled samples of non-hotspots into training and test sets,  with the training set comprising 50\%--85\% of the samples at increments of 5\%.
To evaluate the prediction performance, we compare our
PUMA method with eleven existing and potential approaches described in Section \ref{sec:sim:alter}.
Prediction accuracy is measured by the prediction mean squared error (PMSE), defined as 
$\text{PMSE}=\frac{1}{n_0}\sum_{i=1}^{n_0}(Y_i-\widehat{Y}_i)^2 $, where $n_0$ is the sample size of the test dataset, 
and $Y_i$ and $\widehat{Y}_i$ represent the total number of homeless individuals ($Total$) in the $i$th region and its predicted value, respectively. 
For each sample split ratio, we conduct 500 repeated random splits. In each repetition, we compute the corresponding PMSE and then scale it by subtracting the lowest PMSE among all twelve methods from the original PMSE for ease of presentation. Finally, we report the averaged scaled PMSE along with its standard errors.

Table \ref{scale_PMSE} shows the averaged PMSE under different proportions of training sets.
The results show that the proposed method can consistently yield the most accurate prediction results under different sample splitting ratios, i.e., achieving the lowest PMSE and standard error.
PUMA outperforms all other alternative methods 
because the competing approaches, 
to varying degrees, ignore certain sources of uncertainty.
Among the alternative methods, PML1 delivers the most competitive predictive performance. PML2 performs slightly worse than PML1, while PLAM1 performs substantially worse. This phenomenon can be explained as follows.
First, PML1 outperforms PML2 because the XGBoost used in PML1 provides more accurate predictions than the random forest algorithm in this application. This indicates that ignoring uncertainty in the choice of ML algorithms can lead to suboptimal predictive performance.
Second, PML1 and PML2 outperform PLAM1 because PLAM1 
relies on a fixed power tuning parameter, which makes its performance heavily dependent on the quality of the ML predictions. The lack of adaptivity in tuning prevents PLAM1 from adequately balancing the information from labeled and unlabeled data, resulting in inferior performance. Similar phenomena have been observed in related prediction-powered inference settings 
\citep[e.g.,][]{angelopoulos2023ppi++}.
Other alternative methods based on model selection or those that ignore model uncertainty, including PAIC, PBIC, PLARM, PPI++ML1, PPI++ML2, and LARM, also exhibit inferior predictive performance, underscoring the necessity of accounting for model uncertainty in prediction.
These results further highlight the importance of simultaneously addressing uncertainty arising from model specification, ML algorithms, and power tuning parameters within a unified prediction-powered model averaging framework.
\begin{table}[H]
	\centering{
		\caption{ 
			The averaged PMSEs for the Los Angeles homeless data under different proportions of training sets, with the corresponding standard errors reported in parentheses.
		}
		\label{scale_PMSE}
		\setlength{\tabcolsep}{2mm}{ 
			\begin{tabular}{lccccccccccccccc} 
				\toprule[1.3pt]
				& 50\% & 55\% & 60\% & 65\% & 70\% & 75\% & 80\% & 85\% \\  
				\midrule[1.5pt]
				\multirow{2}{*}{PUMA} & \textbf{0.380} & \textbf{0.359} & \textbf{0.357} &\textbf{0.346}   &\textbf{0.333} &\textbf{0.335}    &\textbf{0.345}  &\textbf{0.371}   \\\addlinespace[4pt]
				& (0.052) & (0.049) & (0.046) & (0.039) & (0.040) & (0.035) & (0.034) & (0.034) \\ \addlinespace[10pt]
				\multirow{2}{*}{PEMA} & 0.475 & 0.430  &0.416  &   0.409  & 0.419  &   0.447  & 0.476  & 0.512    \\\addlinespace[4pt]
				& (0.052) & (0.047) & (0.047) & (0.045) & (0.046) & (0.047) & (0.050) & (0.054) \\ \addlinespace[10pt]
				\multirow{2}{*}{PAIC} & 0.500 & 0.466  &0.450   &    0.422 &  0.391 &  0.368 &     0.368 & 0.388    \\\addlinespace[4pt]
				& (0.066) & (0.063) & (0.061) & (0.052) & (0.053) & (0.044) & (0.045) & (0.047) \\ \addlinespace[10pt]
				\multirow{2}{*}{PBIC} &0.577 & 0.585 &0.573  &    0.564  & 0.538  &     0.532   &    0.526  &0.560     \\\addlinespace[4pt] 
				& (0.057) & (0.056) & (0.052) & (0.045) & (0.047) & (0.044) & (0.045) & (0.051) \\ \addlinespace[10pt]
				\multirow{2}{*}{PLARM} & 0.533& 0.492 &0.469 &   0.452  &   0.431 &    0.431   &    0.442 & 0.471    \\\addlinespace[4pt]
				& (0.060) & (0.055) & (0.053) & (0.046) & (0.045) & (0.041) & (0.041) & (0.043) \\ \addlinespace[10pt]
				\multirow{2}{*}{PML1} &0.385 & 0.365  &0.362 &    0.351 & 0.339 &   0.340   &     0.349  & 0.376    \\\addlinespace[4pt]
				& (0.052) & (0.049) & (0.047) & (0.040) & (0.041) & (0.036) & (0.035) & (0.034) \\ \addlinespace[10pt]
				\multirow{2}{*}{PML2} &0.385 & 0.368  &0.365 &    0.356 &  0.344  &     0.346   &    0.354  & 0.379      \\\addlinespace[4pt]
				& (0.053) & (0.050) & (0.048) & (0.040) & (0.041) & (0.036) & (0.036) & (0.035) \\ \addlinespace[10pt]
				\multirow{2}{*}{PLAM1} & 1.258  & 1.089  &0.996 &     0.916  & 0.890  &    0.895    &    0.890  & 0.913     \\\addlinespace[4pt]
				& (0.118) & (0.106) & (0.102) & (0.098) & (0.100) & (0.102) & (0.106) & (0.117) \\ \addlinespace[10pt]
				\multirow{2}{*}{PLAM0} & 0.419& 0.399 &0.392 &     0.378   &  0.361     &   0.358   &     0.362      &0.386   \\\addlinespace[4pt]
				& (0.059) & (0.056) & (0.053) & (0.045) & (0.046) & (0.041) & (0.040) & (0.041) \\ \addlinespace[10pt]
				\multirow{2}{*}{PPI++ML1} & 0.548 & 0.501 &0.476  &  0.455  &  0.433   &    0.431      & 0.442      &    0.470 \\\addlinespace[4pt]
				& (0.064) & (0.058) & (0.056) & (0.048) & (0.047) & (0.042) & (0.042) & (0.043)   \\ \addlinespace[10pt]
				\multirow{2}{*}{PPI++ML2} & 0.544    &0.501  &0.477  & 0.459 &  0.436 &    0.435 &0.445        &     0.473  \\\addlinespace[4pt]
				& (0.062) & (0.057) & (0.055) & (0.047) & (0.046) & (0.042) & (0.042) & (0.043) \\ \addlinespace[10pt]
				\multirow{2}{*}{LARM} &  0.554   & 0.507 &    0.480  & 0.459  &  0.436  &  0.434  &     0.444   &   0.472 \\\addlinespace[4pt]
				& (0.065) & (0.059) & (0.056) & (0.048) & (0.047) & (0.042) & (0.042) & (0.043) \\
				\bottomrule[1.3pt]
			\end{tabular}
			\begin{tablenotes}
				\footnotesize
				\setlength{\hangindent}{2em}  
				\setlength{\hangafter}{1}    
				\item {The PMSE obtained from each repetition is multiplied by $10^{-2}$.}
			\end{tablenotes}
	} }
\end{table}

\section{Discussion}\label{Discussion}
In this paper, we propose a novel
prediction-powered linear regression framework,
called Prediction-powered Unified Model Averaging (PUMA),
which leverages unlabeled data to
improve predictive accuracy
while preserving model interpretability.
Our approach incorporates model averaging
to systematically address the uncertainty arising
from model specification,
power tuning parameter selection,
and the choice of ML algorithms.
We establish the in-sample and out-of-sample asymptotic optimality, estimation consistency, and asymptotic distribution of the proposed PUMA estimator, and compare
it with several recent and competing methods.
Both extensive simulation studies and a real data application support our theoretical findings.

This work also opens several promising
directions for future research.
First, the convergence rate of the optimal weight vector
remains unclear and warrants further investigation.
Second, integrating large language models into the prediction-powered mechanisms to enable information extraction from unstructured data represents
an exciting frontier.
Developing adaptive weighting strategies
for prediction-powered large language models
is of particular interest
and is left for future work.

\section*{Declaration of Competing Interest}
The authors declare that they have no known competing financial interests or personal relationships that could have appeared to 
influence the work reported in this paper.

\section*{Acknowledgements}
The authors are grateful to Professor Wanfeng Liang for the valuable help provided in the data collection process. This work was supported by the National Key R\&D Program of China under Grant 2024YFA1012200; the National Natural Science Foundation of China under Grant 72525001, 72495124 and 12571285; Beijing Natural Science Foundation under Grant Z240004; and the China Postdoctoral Science Foundation under Grant 2025M773107.

\section*{Supplementary Material}
Supplementary material associated with this paper can be found in the online version.

\bibliographystyle{elsarticle-harv}
\bibliography{bibfile}

\end{document}